\documentclass[preprint]{aastex}
\def\simgr{\,\hbox{\hbox{$ > $}\kern -0.8em \lower 1.0ex\hbox{$\sim$}}\,}
\def\simle{\,\hbox{\hbox{$ < $}\kern -0.8em \lower 1.0ex\hbox{$\sim$}}\,}
\shortauthors{SHEETS and THORSTENSEN.}
\shorttitle{Spectroscopy of HL CMa}
\begin{document}
\title{A Spectroscopic Study of HL Canis Majoris
\footnote{Based on observations obtained at the MDM Observatory, operated by
Dartmouth College, Columbia University, Ohio State University, and
the University of Michigan.}
}

\author{H. A. Sheets and John R. Thorstensen}
\affil{Department of Physics and Astronomy\\
6127 Wilder Laboratory, Dartmouth College\\
Hanover, NH 03755-3528;\\
h.a.sheets@dartmouth.edu}

\begin{abstract}

We present optical spectroscopy of the dwarf nova HL Canis Majoris over
a span of four years.  The observations were made during standstill, outburst,
and quiescence.  We determine an orbital period of $0.2167867 \pm 0.0000017$
days, based on radial velocities determined from H$\alpha$, H$\beta$,
and HeI $\lambda$ 5876 emission.  We also present equivalent widths of the spectral
features in outburst and in quiescence.

\end{abstract}

\section{Introduction}

HL Canis Majoris, despite being one of the brightest dwarf novae, was long overlooked
because of its location in the sky 9 arcminutes south of Sirius.  HL CMa was first detected 
when observations of Sirius by
{\it Einstein} showed a nearby X-ray source, whose location matched that of a variable
star on plates from the Harvard archive \citep{chl81};  subsequent optical 
photometry and spectroscopy exhibited characteristics typical of dwarf novae, such
as recurring outbursts on an approximately 15-day cycle and Balmer emission lines. 
Further photometric study revealed standstills in 1982 and 1999 \citep{99stand} that 
led to the further classification of HL CMa as a Z Cam star.  In 2001, the variable 
also exhibited an unusual outburst cycle of roughly 30 days \citep{kato02}. 

A spectroscopic study of HL CMa by \citet{hutch} gave an orbital period of either 
0.22 or 0.18 d.  They obtained spectra during both outburst and quiescence
and found no conclusive evidence in their data for orbital phase-dependent phenomena.
\citet{wargau} then studied optical spectra of the rise to outburst, maximum, and 
decline from outburst.  They found a period of 0.2145(4) d
\footnote{Uncertainties in the last quoted digits are given in parentheses.} 
and noted absorption 
wings around the Balmer and HeI lines.  
Most recently, \citet{still} have given a period of either 0.2146(4) or 0.2212(5) d,
based on data obtained during outburst.  They also found that the Balmer lines were 
composed of a narrow core, brightest as it shifts from red to blue, and a broader
component that does not vary greatly in brightness. 
  
After more than two decades of study, HL CMa still lacked a precisely determined 
orbital period.  For this reason, we collected spectra over many
runs in order to suppress aliasing and determine the period without ambiguity.  In the 
process of gathering these observations, an improved finding chart was also created 
which is now available from the Living 
Edition of the Catalog and Atlas of Cataclysmic Variables \citep{downes}.

\section{Observations}

Spectra of HL Canis Majoris were taken over 14 observing runs between 2000 January 
and 2004 March from the MDM Observatory at Kitt Peak, Arizona, using mainly the 2.4m 
Hiltner telescope.  The 2001 December data were obtained on the 1.3m McGraw-Hill 
telescope.  The spectra were obtained during both outburst and quiescence.  Two
spectra, taken in 2000 January, were obtained during the standstill that began in 1999
\citep{99stand}, and many spectra were obtained during the unusual outburst state in 
2001-2002 reported by \citet{kato02}. 

The observation and reduction procedures for the Hiltner telescope data were mostly as
described in \citet{tft04}.  In brief, we used the modular
spectrograph and a SITe $2048^{2}$ CCD detector, with 2.0 \AA\ pixel$^{-1}$ from 4300
to 7500 \AA\ and a spectral resolution of 3.5 \AA.  The McGraw-Hill observations 
were made with the Mark III spectrograph and a SITe $1024^{2}$ CCD detector, with 2.2 
\AA\ pixel$^{-1}$ from 4500 to 6800 \AA\ and a spectral resolution of 5.0 \AA.  Many
comparison Hg-Ne-Xe lamp spectra were taken throughout each night in the runs from 
January 2000 through October 2003, while the night sky lines were used for the 
wavelength zero point for the January and March 2004 data. The spectra were then 
reduced using standard IRAF\footnote{IRAF is distributed by the National Optical 
Astronomy Observatories.} procedures. We observed flux standard stars when conditions
appeared clear, and the flux calibration was checked by calculating the $V$ magnitudes 
from the spectra using the passband tabulated by \citet{Bessell}.  The calculated $V$ 
magnitudes for outburst spectra were between 12.5 and 11.6, similar to the 11.7 given 
by the Living Edition \citep{downes}, while those of the quiescent spectra were close to 
14.0, while the Living Edition gives a magnitude of 14.5.  Typically, the flux 
calibrations for our setup are accurate to approximately 10 to 20 percent, with 
systematic errors dominating.

\section{Results}

The quiescent spectra show Balmer line emission, as well as HeII $\lambda 4686$ and 
several HeI lines, all of which are typical of dwarf novae in quiescence. In outburst, 
the spectra show broad absorption wings around H$\beta$, H$\gamma$, and HeI $\lambda 
5876$, as noted in previous studies by \citet{wargau} and \citet{still}.  The CIII/NIII 
feature also stands out strongly just blueward of HeII $\lambda 4686$.  This feature is 
not as clear in the quiescent spectra but seems to be weakly present.  Table 1 gives 
the equivalent widths and FWHM for each state.  Figure 1 shows the average of the 
January 2002 spectra and that of the October 2003 spectra, from which the measurements 
in Table 1 were determined.  The two spectra obtained during standstill, as well as 
those obtained during the unusual outburst cycle, do not appear to differ significantly
from spectra obtained during a normal outburst. 

The radial velocities, given in Table 2, were determined from the H$\alpha$ emission 
line using convolution methods set forth by \citet{schneider} and \citet{shafter}.  
Shafter's technique convolves the spectral line with a positive and a negative Gaussian 
separated by a given width, instead of the derivative of a Gaussian used by Schneider 
and Young.  We used Gaussians each with a width of 6 \AA\ and separated by 14 \AA\ for 
our data.  The velocities had uncertainties estimated from counting statistics errors 
of typically 7 km s$^{-1}$ or less.  We searched for the orbital frequency using the ``residualgram" 
method \citep{tpst96}, which gave a 
well-determined frequency of 4.61282 d$^{-1}$.  The large number of observing runs, 
coupled with long sequences of observations during those runs, resulted in an 
unambiguous cycle count over the four-year span of the data set.  Figure 2 shows the 
periodigram of this search.  The Monte Carlo test described by \citet{tf85} gives a 
discriminatory power of 998/1000 for the 4.61282 d$^{-1}$ frequency, which corresponds 
to an orbital period of 0.216787 d.  The velocities, weighted by their uncertainties,
 were then fit to a sinusoid 
$v(t) = \gamma + K\sin[2\pi(t-T_0)/P]$.  The parameters of the fit are given in Table 3, 
and the folded velocity plot is shown in Figure 3. 

While the period we determined is similar to those suggested in previous studies of 
HL CMa, it still differs from them by more than the margin of error in each.  
%Because our data cover many baselines, it is better 
%suited to long-term period searches, and thus we are confident that our period is correct.
To further ensure that our period is correct, we determined the radial velocities from 
the H$\alpha$ line using the original convolution method of \citet{schneider}, and also 
determined the velocities from the H$\beta$ and HeI $\lambda$ 5876 lines using both 
methods.  The errors in the velocities from H$\beta$ and HeI $\lambda$ 5876 were typically 
of the order of 10 to 20 km s$^{-1}$.  The results from the period searches of each set of 
velocities are listed in Table 3.  The H$\beta$ velocities did not independently confirm 
our choice of cycle count, evidently because of inadqueate signal-to-noise,
but did show strong power at our adopted period; the fits listed in Table 3 are for
the alias which matches our preferred ephemeris.  In general, the less-noisy
fits in Table 3 give periods more closely matching the adopted period.
%When we ran the first set of H$\beta$ velocities through 
%the period-search program, it picked out one of the other aliases as the best fit, with a $\sigma$
%of 43 km s$^{-1}$.  The alias listed in the table from this set of velocities was the third-best alias and was 
%closest to the period of 0.216787 d that we determined from the first set of H$\alpha$ 
%velocities.  For similar reasons, the alias listed for the second set of H$\beta$ velocities is the 
%second-best alias.  As the errors in both the velocities and the fits decrease, the resulting
%period more closely matches that derived from the first set of H$\alpha$ velocities.  
As a final check, we 
split up the first set of H$\alpha$ velocities into two sets according to the state of the system.  
Performing the period search on each of these sets both produced essentially the same period as 
the original search on the full data set.  Thus we are confident that our period is correct.

\section{Conclusion}

We have determined the orbital period of HL CMa unambiguously to be 0.2167867(17) d.  We also have found no 
significant differences between spectra taken during standstill, during outburst, and during the unusual 
outbursting in 2001-2002.  

{\it Acknowledgments.} 
We would like to thank Bill Fenton, who took a portion of the 2.4m data, as well as Cindy Taylor of
the Lawrenceville School, who took the 1.3m data.  We gratefully acknowledge support from NSF grants
AST-9987334 and AST-0307413.  We also thank the MDM staff for their able observing support.

\clearpage

% figure 1
\begin{figure}
\epsscale{1.0}
\plotone{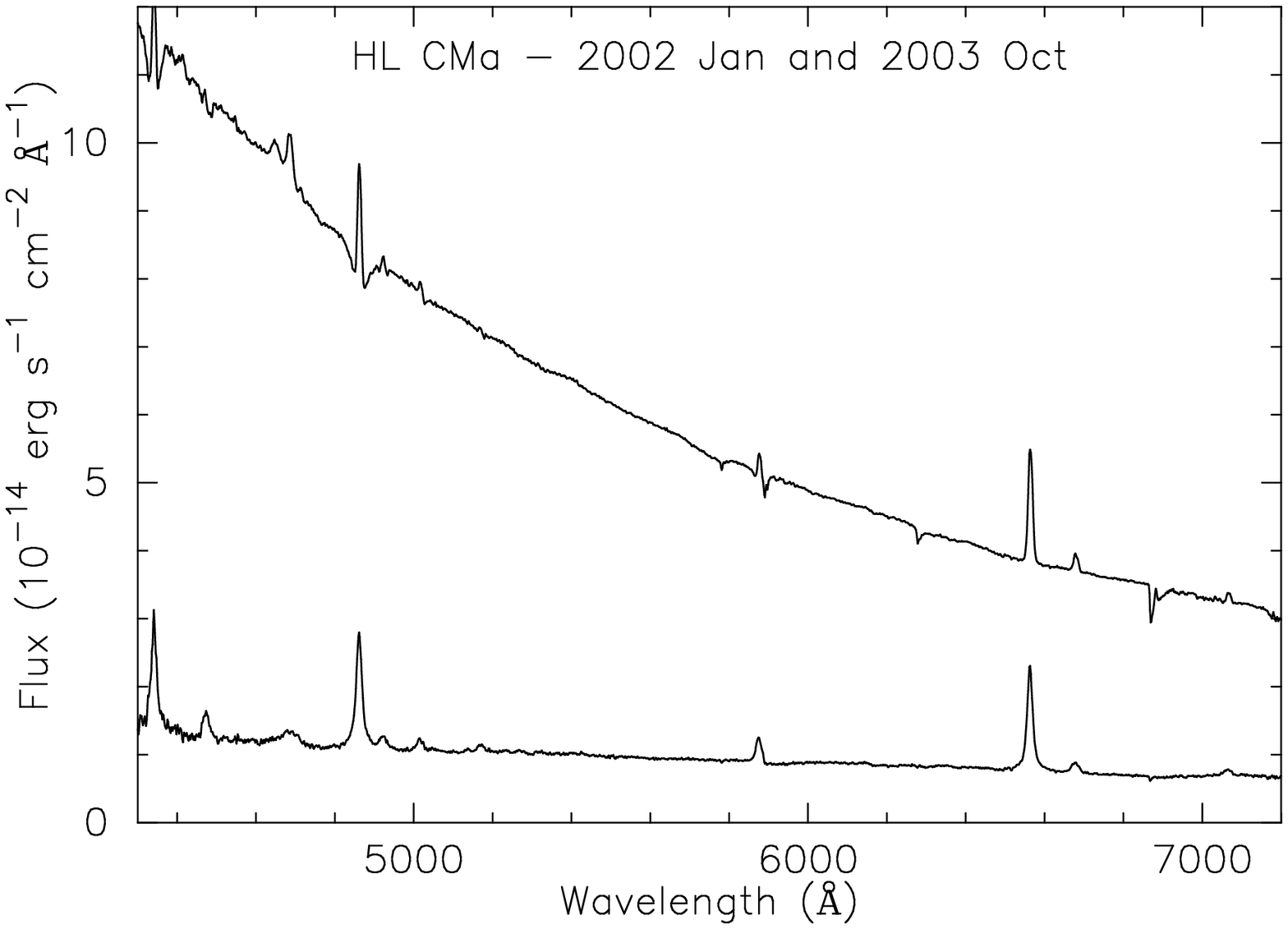}
\caption{The averaged spectra from 2002 January (top) and 2003 October (bottom) are plotted.}
\end{figure}

\clearpage

% figure 2
\begin{figure}
\epsscale{1.0}
\plotone{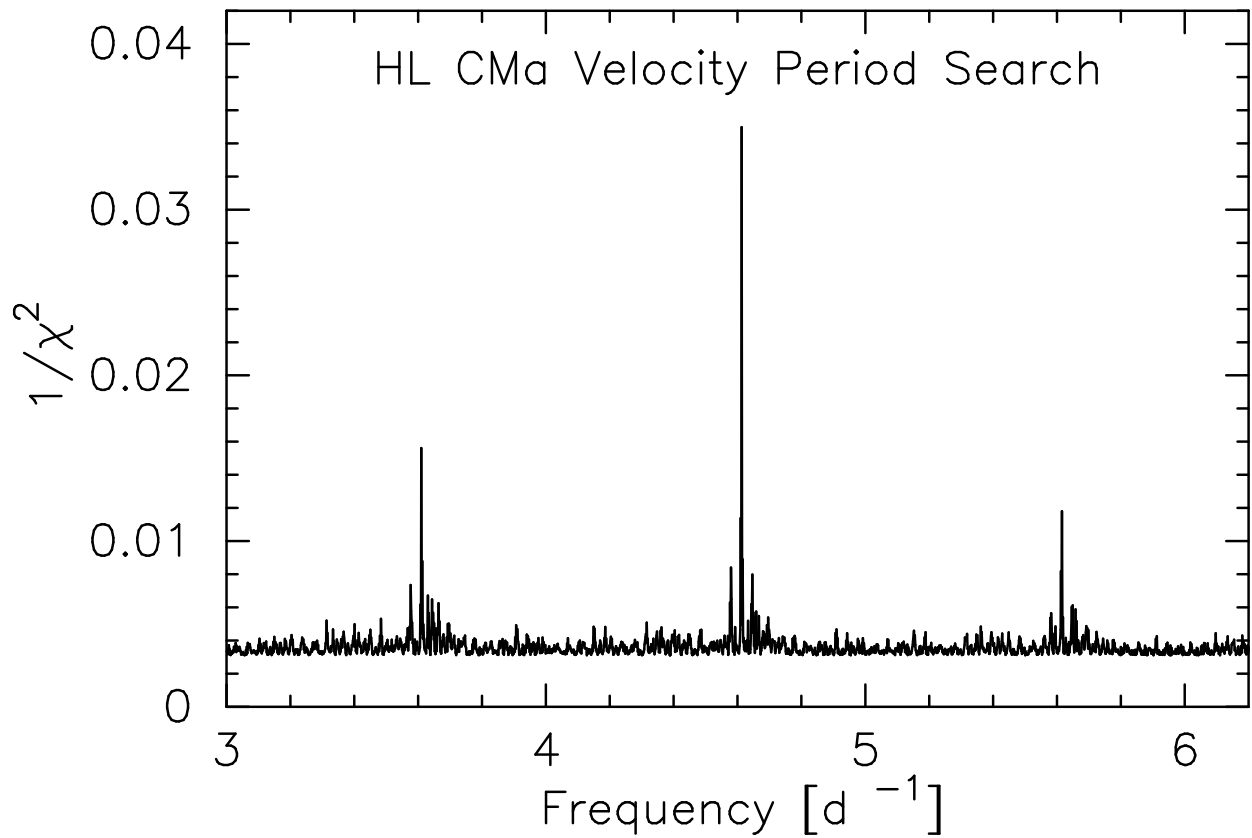}
\caption{The tallest peak, in the center, corresponds with the best fit frequency of 4.61 cycles 
day$^{-1}$. The two next-best fits are the aliases at 3.61 cycles day$^{-1}$ and 5.62 cycles
day$^{-1}$.  Frequency space is oversampled by a factor of 12 in the original search.  This figure 
was created by plotting the local maxima of the original periodigram and connecting those points 
with straight lines.}
\end{figure}

\clearpage

% figure 3
\begin{figure}
\epsscale{1.0}
\plotone{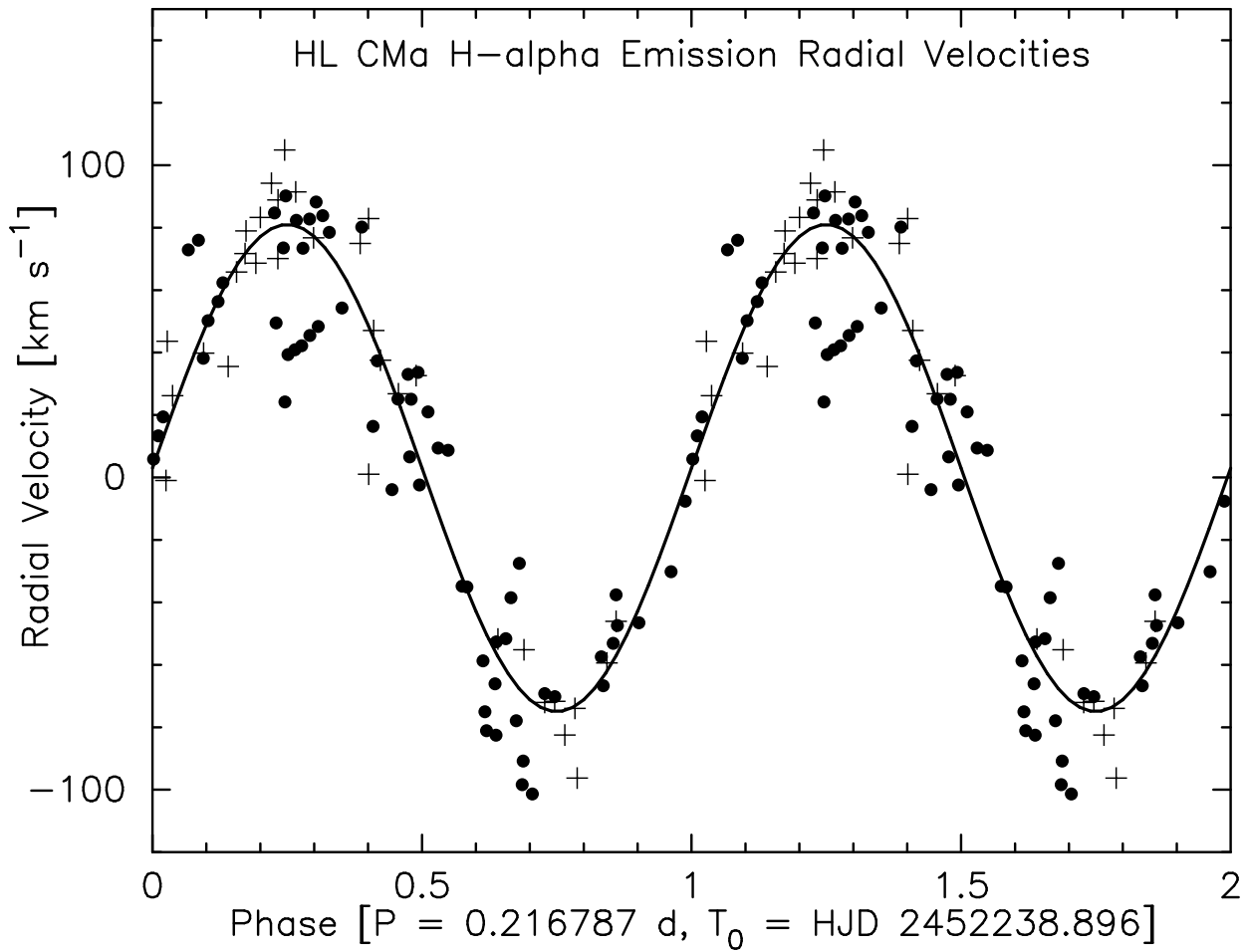}
\caption{Outburst spectra are marked with solid circles, and quiescent spectra are marked with pluses.  The data are
plotted twice for continuity.}
\end{figure}

\clearpage

% table 1
\begin{deluxetable}{lrcc}
\tabletypesize{\scriptsize}
\tablewidth{0pt}
\tablecolumns{4}
\tablecaption{Spectral Features in Quiescence and Outburst}
\tablehead{
&
\colhead{E.W.\tablenotemark{a}} &
\colhead{Flux}  &
\colhead{FWHM \tablenotemark{b}} \\
\colhead{Feature} &
\colhead{(\AA )} &
\colhead{(10$^{-16}$ erg cm$^{-2}$ s$^{1}$)} &
\colhead{(\AA)} \\
}
\startdata
\cutinhead{Quiescence - 2003 Oct:}
           H$\gamma$ & $17.7$ & $2600$ & 16 \\
  HeI $\lambda 4471$ & $ 5.4$ & $ 680$ & 19 \\
 HeII $\lambda 4686\tablenotemark{c}$ & $8.8$ & $1000$ & \nodata \\
            H$\beta$ & $29.2$ & $3400$ & 19 \\
  HeI $\lambda 4921$ & $ 2.3$ & $ 260$ & 20 \\
  HeI $\lambda 5015$ & $2.5$ & $230$ & \nodata \\
   Fe $\lambda 5169$ & $ 2.0$ & $ 210$ & \nodata \\
  HeI $\lambda 5876$ & $ 6.6$ & $ 600$ & 16 \\
           H$\alpha$ & $37.1$ & $3000$ & 18 \\
  HeI $\lambda 6678$ & $ 4.1$ & $ 300$ & 21 \\
\cutinhead{Outburst - 2002 Jan:}
           H$\gamma$ & $-1.5$\tablenotemark{d} & $-1800$ & 10\tablenotemark{e} \\
  HeI $\lambda 4471$ & $ 0.1$ & $120$ &  7 \\
CIII/NIII & $0.5$ & $450$ & \nodata \\
 HeII $\lambda 4686$ & $0.8$ & $790$ & 14 \\
            H$\beta$ & $-0.3$\tablenotemark{d} & $-290$ & 10\tablenotemark{e} \\
  HeI $\lambda 4921$ & $0.3$ & $240$ &  9 \\
  HeI $\lambda 5015$ & $0.3$ & $200$ &  9 \\
   Fe $\lambda 5169$ & $0.1$ & $ 46$ &  7 \\
  HeI $\lambda 5876$ & $-0.5$\tablenotemark{f} & $-240$ &  9\tablenotemark{e} \\
           H$\alpha$ & $6.0$ & $2300$ & 13 \\
  HeI $\lambda 6678$ & $1.0$ & $350$ & 13 \\
\enddata

\tablenotetext{a}{Emission equivalent widths are counted as positive.}
\tablenotetext{b}{From Gaussian fits.}
\tablenotetext{c}{Includes weak CIII/NIII blend.}
\tablenotetext{d}{Includes the absorption wings.}
\tablenotetext{e}{Fit to the emission core of the feature.}
\tablenotetext{f}{Includes the asborption wings as well as interstellar Na D absorption,
        indistinguishable from the wings.}

\end{deluxetable}

\clearpage

% table 2
\begin{deluxetable}{lcclcclcc}
\tabletypesize{\scriptsize}
\tablewidth{0pt}
\tablecolumns{9}
\tablecaption{Radial Velocities}
\tablehead{
\colhead{Time\tablenotemark{a}} &
\colhead{$v_{\rm emn}$} &
\colhead{HA} &
\colhead{Time\tablenotemark{a}} &
\colhead{$v_{\rm emn}$} &
\colhead{HA} &
\colhead{Time\tablenotemark{a}} &
\colhead{$v_{\rm emn}$} &
\colhead{HA} \\
 &
\colhead{(km s$^{-1}$)} &
 &
 &
\colhead{(km s$^{-1}$)} &
 &
 &
\colhead{(km s$^{-1}$)} &
 }
\startdata
51549.8790 & $-98$ & $+1:49$ &52237.8310 & $76$ & $-2:06$ & 52295.6970 & $13$ & $-1:33$ \\
51549.8830 & $-101$ & $+1:55$ &52237.9460 & $-81$ & $+0:40$ & 52295.6980 & $19$ & $-1:31$ \\
51641.6120 & $-67$ & $+1:33$ &52237.9500 & $-66$ & $+0:45$ & 52295.8810 & $-47$ & $+2:53$ \\
51641.6160 & $-53$ & $+1:38$ &52238.8600 & $-57$ & $-1:21$ & 52296.8270 & $85$ & $+1:39$ \\
51642.6140 & $25$ & $+1:40$ &52238.8660 & $-38$ & $-1:12$ & 52297.6460 & $6$ & $-2:38$ \\
51642.6180 & $33$ & $+1:45$ &52262.8310 & $16$ & $-0:29$ & 52297.7830 & $-83$ & $+0:40$ \\
51642.6220 & $34$ & $+1:51$ &52262.8390 & $-4$ & $-0:18$ & 52324.6850 & $-69$ & $+0:06$ \\
51642.6260 & $21$ & $+1:57$ &52262.8460 & $7$ & $-0:08$ & 52324.6890 & $-70$ & $+0:12$ \\
51642.6300 & $9$ & $+2:03$ &52268.6880 & $38$ & $-3:32$ & 52324.7930 & $49$ & $+2:42$ \\
51642.6340 & $9$ & $+2:08$ &52268.6950 & $27$ & $-3:22$ & 52324.7980 & $39$ & $+2:49$ \\
51644.6100 & $-38$ & $+1:42$ &52268.7020 & $33$ & $-3:12$ & 52325.6640 & $24$ & $-0:20$ \\
51644.6140 & $-28$ & $+1:48$ &52270.8070 & $83$ & $-0:32$ & 52325.6680 & $41$ & $-0:14$ \\
51645.6100 & $42$ & $+1:46$ &52270.8140 & $89$ & $-0:22$ & 52325.7490 & $-53$ & $+1:43$ \\
51645.6130 & $45$ & $+1:50$ &52270.8210 & $91$ & $-0:12$ & 52325.7530 & $-52$ & $+1:48$ \\
51645.6170 & $48$ & $+1:56$ &52270.8280 & $77$ & $-0:02$ & 52326.6120 & $-75$ & $-1:31$ \\
51843.9160 & $-1$ & $-1:54$ &52293.6300 & $25$ & $-3:17$ & 52327.6860 & $-35$ & $+0:20$ \\
51843.9180 & $26$ & $-1:52$ &52293.6340 & $-2$ & $-3:11$ & 52327.6880 & $-35$ & $+0:23$ \\
51992.6790 & $73$ & $+2:12$ &52293.7410 & $-8$ & $-0:37$ & 52574.0280 & $-47$ & $+0:45$ \\
51994.6250 & $94$ & $+1:02$ &52293.8190 & $54$ & $+1:16$ & 52577.0160 & $-55$ & $+0:40$ \\
51994.6300 & $105$ & $+1:09$ &52293.8920 & $-91$ & $+3:01$ & 52620.8660 & $-30$ & $-0:07$ \\
51995.6150 & $-96$ & $+0:51$ &52294.6310 & $38$ & $-3:12$ & 52620.9590 & $80$ & $+2:07$ \\
52233.8510 & $-72$ & $-1:53$ &52294.6330 & $50$ & $-3:09$ & 52621.8880 & $-78$ & $+0:29$ \\
52233.8550 & $-72$ & $-1:47$ &52294.6370 & $56$ & $-3:03$ & 52624.8140 & $79$ & $-1:06$ \\
52233.8590 & $-82$ & $-1:41$ &52294.6390 & $62$ & $-3:00$ & 52671.7360 & $-59$ & $+0:07$ \\
52233.8630 & $-74$ & $-1:36$ &52294.7010 & $37$ & $-1:31$ & 52927.0190 & $69$ & $-0:15$ \\
52233.9970 & $1$ & $+1:38$ &52294.8810 & $90$ & $+2:49$ & 52927.9840 & $-52$ & $-1:01$ \\
52234.8610 & $75$ & $-1:35$ &52294.8850 & $82$ & $+2:55$ & 52929.0180 & $47$ & $-0:08$ \\
52234.8640 & $83$ & $-1:30$ &52294.8880 & $73$ & $+2:59$ & 53017.8420 & $36$ & $+1:23$ \\
52234.9600 & $-59$ & $+0:48$ &52294.8900 & $83$ & $+3:02$ & 53017.8460 & $66$ & $+1:29$ \\
52234.9630 & $-46$ & $+0:53$ &52294.8930 & $88$ & $+3:06$ & 53018.7160 & $72$ & $-1:35$ \\
52235.0000 & $44$ & $+1:46$ &52294.8950 & $84$ & $+3:09$ & 53018.7290 & $70$ & $-1:16$ \\
52237.8270 & $73$ & $-2:12$ &52294.8980 & $79$ & $+3:14$ & 53066.6090 & $40$ & $-0:57$ \\
\enddata
\tablenotetext{a}{Heliocentric Julian data of
mid-exposure, minus 2 400 000.}
\end{deluxetable}

\clearpage

% table 3
\begin{deluxetable}{lllrrcc}
\tablecolumns{7}
\tabletypesize{\scriptsize}
\tablewidth{0pt}
\tablecaption{Fit to Radial Velocities}
\tablehead{
\colhead{Data set} &
\colhead{$T_0$\tablenotemark{a}} &
\colhead{$P$} &
\colhead{$K$} &
\colhead{$\gamma$} &
\colhead{$N$} &
\colhead{$\sigma$\tablenotemark{b}}  \\
&
&
\colhead{(d)} &
\colhead{(km s$^{-1}$)} &
\colhead{(km s$^{-1}$)} &
&
\colhead{(km s$^{-1}$)} \\
}
\startdata
H$\alpha$\tablenotemark{c} 	& 52238.896(4) & 0.216787(3) &  78(7) & 3(6) & 96 & 19 \\
H$\alpha$\tablenotemark{d} 	& 52233.912(5) & 0.216787(4) & 77(10) & 7(7) & 94 & 24 \\
H$\beta$\tablenotemark{e}  	& 52233.89(1) & 0.216795(12) & 64(17) & 2(13) & 82 & 40 \\
H$\beta$\tablenotemark{f}  	& 52233.893(6) & 0.216791(8) & 66(12) & 2(8) & 88 & 38 \\
H$\beta$\tablenotemark{g}	& 52233.895(8) & 0.216787(9) & 71(14) & 2(10) & 93 & 34 \\
HeI $\lambda 5876$\tablenotemark{h} & 52238.867(7) & 0.216786(5) & 74(14) & 7(10) & 73 & 46 \\
HeI $\lambda 5876$\tablenotemark{i} & 52233.885(5) & 0.216782(4) & 87(10) & 8(8) & 87 & 37 \\
Weighted Average		& \nodata	& 0.2167866(17)	& \nodata & \nodata & \nodata \\
\enddata

\tablenotetext{a}{Heliocentric Julian Date minus 2400000.  The epoch is chosen
to be near the center of the time interval covered by the data, and
within one cycle of an actual observation.}
\tablenotetext{b}{Root-mean-square residual of the fit.}
\tablenotetext{c}{Velocities determined using two gaussians each 3 pixels wide and separated by 7 pixels.}
\tablenotetext{d}{Velocities determined using a derivative of a gaussian 7 pixels wide.}
\tablenotetext{e}{3rd best period, fit to velocities determined using a derivative of a gaussian 4 pixels wide.}
\tablenotetext{f}{2nd best period, fit to velocites determined using a derivative of a gaussian 4.5 pixels wide.}
\tablenotetext{g}{Velocities determined using two gaussians each 2 pixels wide and separated by 4 pixels.}
\tablenotetext{h}{Velocities determined using a derivative of a gaussian 4.6 pixels wide.}
\tablenotetext{i}{Velocities determined using two gaussians each 2 pixels wide and separated by 4.7 pixels.}
\end{deluxetable}

\end{document}